# Sequential Flipping: A Donor-Acceptor Exchange Mechanism in Water Trimer

Xinrui Yang [a], Rui Liu [a], Ruiqi Xu [a], Zhigang Wang *, [a, b]

[a] Institute of Atomic and Molecular Physics, Jilin University, Changchun 130012, China

[b] International Center for Computational Method & Software, College of Physics, Jilin University, Changchun 130012, China.

*Author to whom correspondence should be addressed: wangzg@jlu.edu.cn

**Abstract**

The donor-acceptor exchange (DAE) is a significant hydrogen bond network rearrangement (HBNR) mechanism because it can lead to the change of hydrogen bond direction. In this work, we report a new DAE mechanism found in water trimer that is realized by sequential flipping (SF) of all molecules rather than the well-known proton transfer (PT) process. Meanwhile, the SF process has a much smaller potential barrier (0.262 eV) than the previously predicted collective rotation process (about 1.7 eV), implying that SF process is a main flipping process that can lead to DAE. Importantly, high-precision ab initio calculations show that SF-DAE can make the water ring to show a clear chiral difference from PT-DAE, which brings the prospect of distinguishing the two confusing processes based on circular dichroism spectra. The reaction rate analysis including the quantum tunneling indicates an obvious temperature-dependent competitive relationship between SF and PT processes, specifically, the SF process dominates above 65 K, while the PT process dominates below 65 K. Therefore, in most cases, the contribution for DAE mainly comes from the flipping process, rather than the PT process as previously thought. Our work enriches the understanding of the DAE mechanism in water trimer and provides a piece of the jigsaw that has been sought to the HBNR mechanism.

**Introduction**

The hydrogen bond network rearrangement (HBNR) underlies various exotic properties of water, and its mechanism receives immense interest.[1,2] As the smallest water cluster with a complete hydrogen bond (H-bond) network, cyclic water trimer $(H_2O)_3$ has long been considered as the key to understand the HBNR.[3–7] However, due to the limitation of experimental resolution in the past, it is difficult to observe all reaction processes, especially those corresponding to a small splitting (< 1 MHz) in vibration-rotation-tunneling (VRT) spectra.[4,8–10] As a result, the previous studies on water trimer usually focused on the simple process, which may lead to omission and even misunderstanding of HBNR.

With such a potential risk, a large number of previous studies have confirmed that the paths of HBNR in water trimer can be divided into flipping and proton transfer (PT) processes. On the one hand, theoretical studies showed that there are two general flipping pathways which are called as single flipping and bifurcation, and the combinations of these can explain the experimentally observed quartets.[7–15] On the other hand, the previous studies confirmed that there is a synergistic transfer pathway of three protons in water trimer that leads to the donor-acceptor exchange (DAE), which is the most fundamental HBNR mechanism.[16,17] The DAE mechanism, also known as the clockwise to counterclockwise (cw-ccw) motion because of the head-to-tail feature of the cyclic water ring, is firstly used to name the low-barrier transition process of water trimer speculated by Pugliano and Saykally in 1992.[16] Due to the breaking of covalent bonds, PT process was generally recognized as the high-barrier path and Saykally's conjecture is believed to be the flipping process. As early as 1993, Wales has been tried to verify whether flipping could lead to DAE in water trimer.[11] But limited by the reliability of the theoretical method and the calculation accuracy at that time, this issue has not been clearly understood. In the following decade, although a large number of calculations based on transition state theory (TST) were used to explore the HBNR processes in water trimer, the cw-ccw path speculated in 1992 which was thought as each water molecule concerted rotating about $C_2$ axis, was not discovered.[12,18,19] Until 2011, such a cw-ccw path was predicted by the ring-polymer instanton (RPI) method extended by elementary graph theory, and authors claimed that the action of this process was very large that the influence of it was thought negligibly small.[20] Anyway, whether there is a low-barrier flipping in water trimer that can lead to DAE, as conjectured by Pugliano and Saykally in 1992, is still puzzling and challenging.

Despite the advances in experimental techniques and high-precision computational methods in recent years, the theoretical exploration of the complete HBNR mechanism in water trimer has instead not been carried out.[4,8–10] Since a molecular jump mechanism based on the flipping of water molecules in water clusters was proposed in 2006, the role of flipping in the HBNR has become an important topic.[3,21,22] Therefore, there is an urgent need for a clear understanding of the DAE mechanism and the basic question that whether the low-barrier flipping can lead to DAE in water trimer requires reconsideration. Additionally, with the discovery of quantum effects, especially quantum tunneling effects in light nuclear elements, it brings an essential perspective for understanding the microscopic mechanism, which will provide an opportunity to revisit and even deepen the understanding of the HBNR mechanism in water trimer.[23–28]

It is with these above considerations that we report a newly discovered DAE mechanism in water trimer through the sequential flipping (SF) process of three water molecules rather than the well-known PT process. Also, we compared and discussed this SF process with reported results in the past, and also carefully discuss the collective rotation process that has been of interest in the past. By comparing the SF and PT processes, there are differences in vibration spectroscopy based on isotope effect and in the chirality of the products of the two processes, which provides promising avenues for further theoretical studies and even experimental observation. Moreover, reaction rate analysis indicates that the quantum tunneling effect leads to the dominance of SF process above 65 K. Our work provides new insights and perspectives on the HBNR mechanisms in water trimer.

## Methods

All structures and spectra were obtained at the MP2/aug-cc-pVTZ level, in which the energies were further corrected by CCSD(T)/aug-cc-pVQZ level.[29,30] While the transition states (TSs) were fully searched, and the potential energy surfaces (PESs) were verified by the intrinsic reaction coordinate (IRC) method.[31,32] All the above calculations were performed using the Gaussian 09 package.[33] The Potential Energy Distribution (PED) analysis of theoretical vibrational spectra was based on the VEDA program.[34] Reaction rate results were obtained from kSC program.[35]

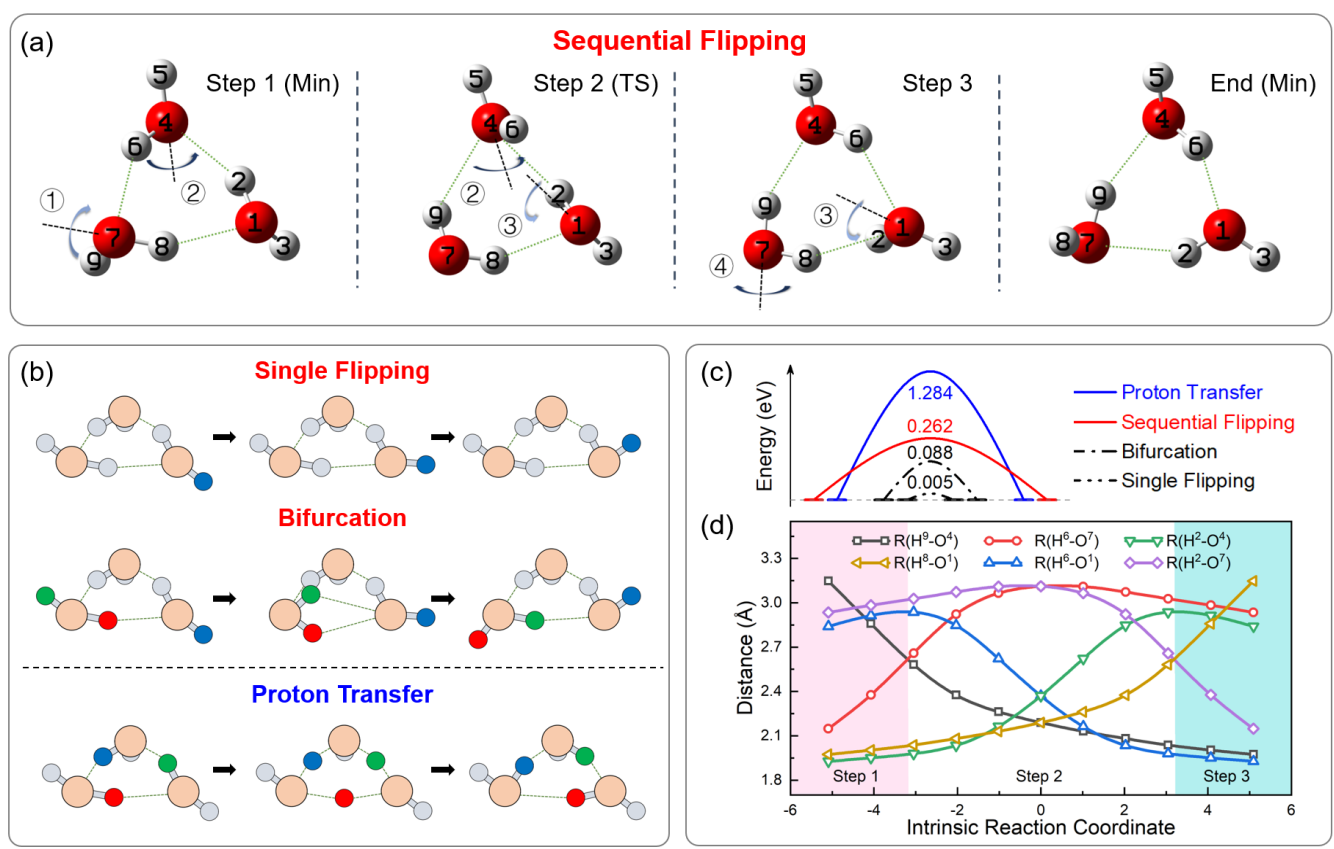

**FIG. 1.** (a) The sequential flipping (SF) process of water trimer. The dotted lines represent the OH axis ($C_2$ axis). (b) Three reported pathways of hydrogen bonding network rearrangement in water trimer. (c) Schematic diagram of energy barriers of four hydrogen bond network rearrangement processes. The electron energies were corrected by CCSD(T)/aug-cc-pVQZ level. (d) The changes of distance between O and H atoms (R(O-H)) during the SF process along the intrinsic reaction coordinate. When x = 0, the structure represents the transition state of SF process.

## Results and discussion

In order to comprehensively understand the HBNR mechanism in water trimer, the reliable reaction paths by global search were obtained based on high-precision ab initio calculations as shown in Fig. 1. Except the previously known flipping and PT processes, we found that there is a hitherto unexplored sequential flipping (SF) process of the whole water ring, which is different with the flipping of only two water molecules or even one water molecule. The calculated energy barriers in Fig. 1(c) indicate that both the single flipping (0.005 eV) and bifurcation (0.088 eV) processes are low-barrier processes, in which the barrier of the single flipping process is even overwhelmed by the zero-point energy as reported before.[36] Compared to the above two general reported flipping processes, the SF process exhibits the highest potential barrier (0.262 eV), but is still much lower than that of the PT process (1.284 eV).

Specifically, the SF process can be regarded as a process in which H atoms rotate around O-H axis in turn and can be divided into three steps. Firstly, the $H^9$-$O^7$ bond rotates around the $H^8$-$O^7$ axis, accompanied by the $H^6$-$O^4$ bond rotating around the $H^5$-$O^4$ axis. This process is accompanied by the breaking of the $H^6\cdots O^7$ H-bond and the generation of the $O^4\cdots H^9$ H-bond. Then, the $H^6$-$O^4$ bond continues to rotate, pushing the $H^2$-$O^1$ bond to rotate along the $H^3$-$O^1$ axis in Step 2 and further pushing the $H^8$-$O^7$ bond to rotate with the $H^9$-$O^7$ axis in Step 3. This character can be also recognized by the variation of the distance between O and rotating H atoms (R(H-O)) as shown in Fig. 1(d). In Step 1, the change of R($H^9$-$O^4$) mainly leads to change of R($H^6$-$O^7$). In Step 2, the change of R($H^6$-$O^1$) leads to changes of R($H^2$-$O^7$) and R($H^2$-$O^4$). Meanwhile, the change of R($H^2$-$O^7$) leads to the change of R($H^8$-$O^1$) in Step 3.

**Table I.** Transition states (TSs) of the sequential flipping process, imaginary frequencies and energies under different basis sets. All calculations are based on the MP2.

| Basis set | TS | ν (cm$^{-1}$) | E (Hartree) |
|---|---|---|---|
| STO-3G | "ts4" in ref. [11] | -44.84 | -225.035 |
| 4-31G | SF | -185.40 | -228.140 |
| 4-31G* | SF | -181.93 | -228.401 |
| DZP | SF | -141.87 | -228.743 |
| 6-31+G | SF | -198.95 | -228.406 |
| 6-31+G(d) | SF | -153.23 | -228.650 |
| def2-TZVP | SF | -147.59 | -228.957 |
| aug-cc-pVTZ | SF | -139.08 | -229.003 |

The reason why this SF process is special is that it is the flipping pathway leading to DAE. The rotation process that can lead to DAE has been a matter of great interest in the field of water trimers. Throughout the history, the similar attempt has been tried to explore in 1993.[11] Theoretical calculations at MP/STO-3G level obtained transition state "ts4" in the article, corresponding to a potential DAE process. However, it's reactant "min2" connected by this reaction do not exist stably. Meanwhile, the "ts4" is irreducible at larger basis sets (4-31G, 4-31G*, DZP), indicating that this reaction pathway is not recognized. Here, we used MP2 method to recalculate the TS of SF process by using different basis sets. As illustrated in Table I, the TSs of the SF process are obtained using different basis sets with similar energies, except for the minimal basis set STO-3G. Interestingly, the result of STO-3G exhibit the same configuration of "ts4" as in the past.[11] However, when using a larger basis set, the TS of the SF process is available again. It explains the fact that this SF process remains undiscovered due to prior limitations of computational resources and computational accuracy.

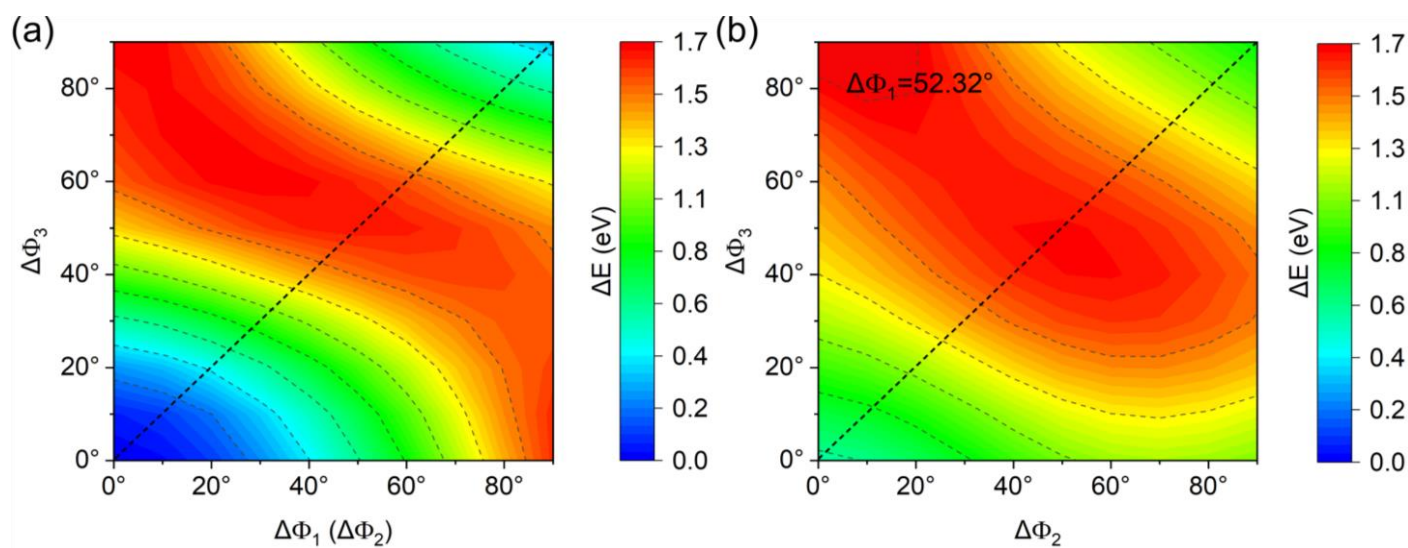

**FIG. 2.** The potential energy surface (PES) of three variables $\Delta\Phi_1$, $\Delta\Phi_2$ and $\Delta\Phi_3$. (a) denotes the PES when $H^2$-$O^1$-$H^3$ and $H^5$-$O^4$-$H^6$ are rotating simultaneously. (b) denotes the PES of $\Delta\Phi_2$ and $\Delta\Phi_3$ when $\Delta\Phi_1$ is fixed to 52.32°, which is the middle value of change during DAE process. Here,

ΔΦ represents the angle of rotation around the $C_2$ axis of each water molecule. Specifically, $ΔΦ_1$ and $ΔΦ_2$ represent water molecules $H^2$-$O^1$-$H^3$ and $H^5$-$O^4$-$H^6$, while $ΔΦ_3$ represents water molecule $H^8$-$O^7$-$H^9$.

In 2018, Li and co-workers reported two kinds of rotational processes leading to DAE in water tetramers on NaCl(001), which are called collective rotation pathway and stepwise rotation pathway, respectively.[23,37] The collective rotation process with smaller barrier is considered to be the simultaneous rotation of four water molecules along the $C_2$ axis, accompanied by the simultaneous breakage and formation of multiple hydrogen bonds. In addition, the SF process is similar to the stepwise rotation pathway. Due to the similarity of collective rotation pathway and stepwise rotation pathway in water tetramer, it would be of interest to consider the collective rotation pathway in water trimer.

In fact, such a flipping process of each water molecule rotating around its $C_2$ axis was expected to be found when the HBNR in water trimers began to be concerned.[16] This process has to be a collective rotation process, because the early flipping of one of the water molecules will certainly hinder the flipping of the next one. Subsequently, this process was formally reported in 2011 named cwccw process and was used as a new result obtained by the ring-polymer instanton (RPI) method extended by elementary graph theory.[20] Before that time, despite the widespread interest, such process has not been reported based on TST.[11,12] We have also made numerous attempts, but have not found a TS corresponding to cw-ccw process. To better understand such rotation processes, we calculated simplified PES with three main variables of rotational angles based on the reported dynamic diagram of the cwccw process.[20] As shown in Fig. 2, when $ΔΦ_1 = ΔΦ_2$, there will be no energy saddle point near the concerted change path of $ΔΦ_1$ and $ΔΦ_3$. When $ΔΦ_1 = 52.32°$, there is only one peak of energy. It is worth noting that the value of the peak is about 1.7 eV, which is much larger than these of SF (0.262 eV) and PT (1.284 eV) processes. This result is similar to the conclusion in the water tetramer that the barrier of stepwise rotation pathway ($>$ 1.2 eV) is higher than that of stepwise rotation (~ 0.7 eV) and PT (~ 0.8 eV) processes.[37] Similar results were also found for reorientation processes of the one-dimensional (1D) water chain.[38] Considering that our method is also reliable for reaction studies, it seems to imply that it is difficult to achieve such processes. In the following discussion, this collective rotation process in water trimer will not be discussed, either because it cannot be verified by TST or has a higher potential barrier.

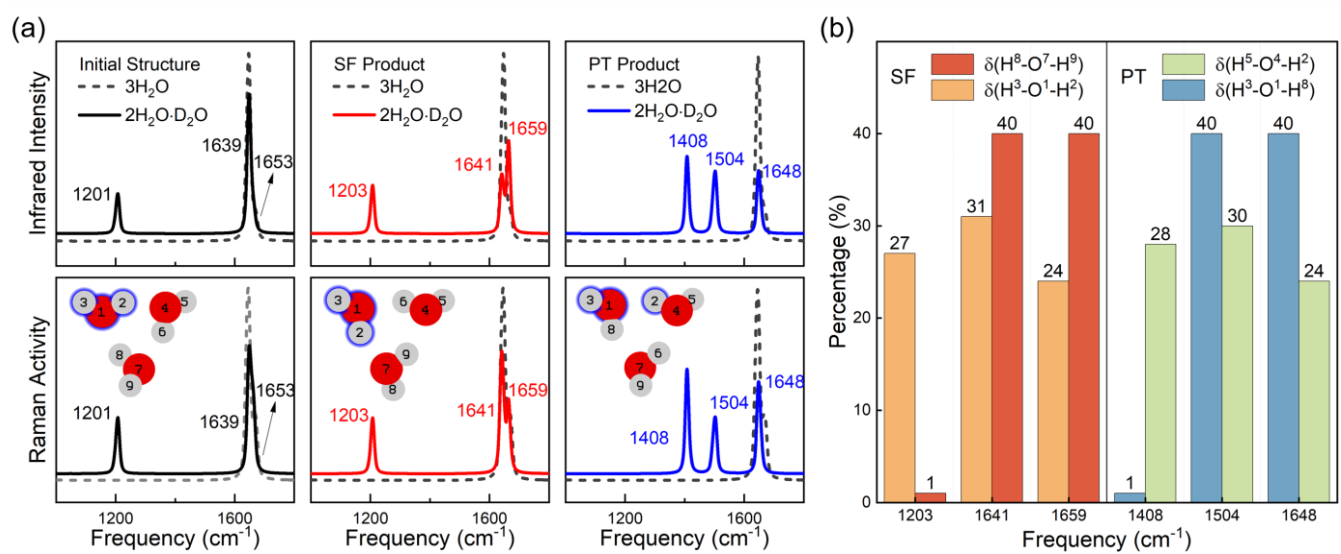

**FIG. 3.** (a) Infrared and Raman spectra of the initial structure, sequential flipping (SF) product and proton transfer (PT) product with one $H_2O$ replaced to $D_2O$. The black lines represent the same initial structure before SF and PT process. The red and blue lines represent the products of SF and PT processes, respectively. The dashed line represents the case with no isotope substitution. (b) The potential energy distribution analysis of the SF and PT processes.

Until then, it was assumed that the PT process was the main DAE mechanism in water trimer. With the discovery of SF-DAE mechanism, the problem of how to distinguish both different PT and SF processes with the similar features, has arisen. Even for future verification, it is necessary to provide a detailed theoretical discussion here.

As shown in Fig. 1(a-b), it can be seen that SF process does not occur the breakage of O-H bonds and the exchange of H atoms whereas the PT process occurs the breakage of O-H bonds and the exchange of H atoms. To clearly describe this difference between the two processes, we studied the Infrared (IR) and Raman spectra considering isotope substitution, as shown in Fig. 3(a). When one $H_2O$ in the initial structure is substituted by one $D_2O$, it will exhibit a significant shifted peak at 1201 cm$^{-1}$. For the SF process, since no exchange of H atoms occurs, but only a conformational shift, it does not change significantly the spectral signatures. In contrast, the PT process caused distinct changes. Due to the exchange of a D atom, the PT process distinguishes the initial structure from the product. Specifically, two D atoms in the product are distributed in two water molecules $H^3$-$O^1$-$H^8$ and $H^5$-$O^4$-$H^2$. It is clear that this change causes the appearance of significant red-shift (from 1201 to 1408 cm$^-$

[1]) and blue-shift (from 1639 to 1504 cm[-1]). These results clearly show the difference between the two processes in terms of whether the O-H bond is broken or not, which will provide an ideal discrimination strategy for future experiments.

To gain a deeper nature of red-shift and blue-shift, we further performed the potential energy distribution (PED) analysis. As shown in Fig. 3(b), the three main moving peaks were assigned, and it can be seen that these vibrational peaks mainly originate from the vibration of the bond angles of water molecules. Specifically, for the product of the SF process, the vibrational peak at 1203 cm[-1] mainly originates from $\delta(H^8\text{-}O^7\text{-}H^9)$ with a 27% percentage. Meanwhile, the peaks at 1641 and 1659 cm[-1] originate from the co-action of $\delta(H^8\text{-}O^7\text{-}H^9)$ and $\delta(H^3\text{-}O^1\text{-}H^2)$ with ratios of 31%, 40% and 24%, 40%, respectively. For the products of the PT process, the main source of the vibrational peak at 1408 cm[-1] is $\delta(H^5\text{-}O^4\text{-}H^2)$, with a 28% contribution. The vibrational peaks at 1504 and 1648 cm[-1] both originate from $\delta(H^5\text{-}O^4\text{-}H^2)$ and $\delta(H^3\text{-}O^1\text{-}H^8)$ with the percentages of 30%, 40% and 24%, 40%, respectively. In addition, it can be seen that the source of the three peaks in PT product contains $\delta(H^5\text{-}O^4\text{-}H^2)$ and $\delta(H^3\text{-}O^1\text{-}H^8)$, which explains the redshift and blueshift caused by the PT process.

From Fig. 3(b), it can be seen that the vibrational sources of the SF and PT processes are different. To clearly show the cause of this phenomenon, the electronic circular dichroism (ECD) and vibrational circular dichroism (VCD) spectra of the products after SF and PT processes were obtained as shown in Fig. 4(a).[28] It is obvious that the peaks of the two processes are clearly symmetric each other, which indicates that there is chiral difference in the two products and the two processes do difference effect on the structure of the water trimer. The PT process is considered to be the process that can transform the chiral isomers, which means that the SF process does not change the chirality of the water ring. Thus, the circular dichroism spectra provide an optional way to experimentally discriminate between the two DAE mechanisms.

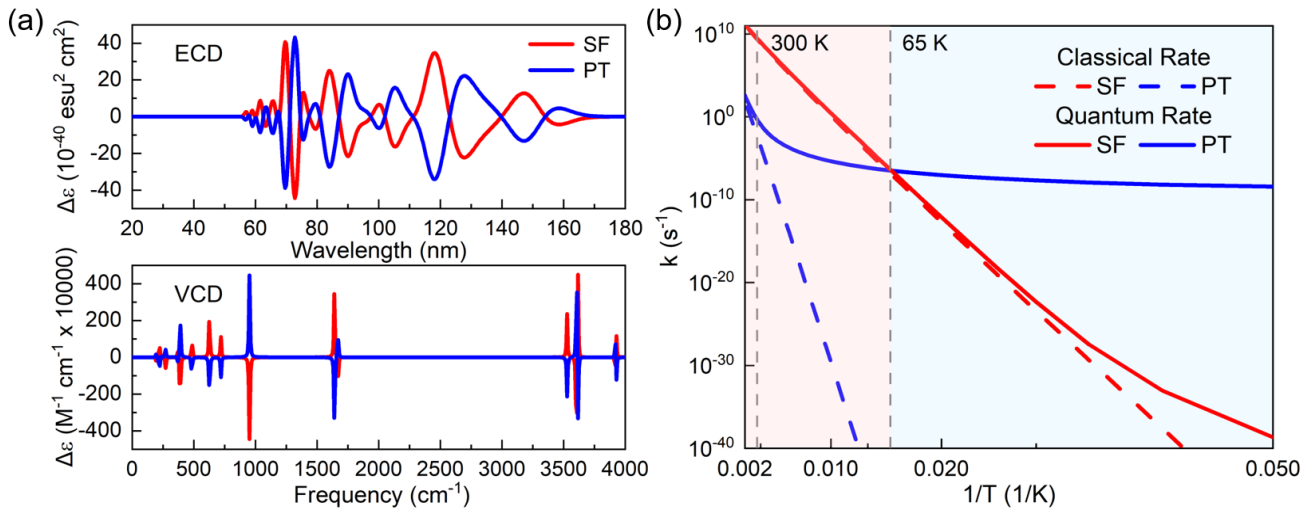

**FIG. 4.** (a) ECD and VCD spectra of the products of sequential flipping (SF) and proton transfer (PT) processes. The red lines and represent SF process and blue lines represent PT process, respectively. (b) Reaction rates for sequential flipping and proton transfer processes. The solid lines represent the quantum-corrected reaction rate, in which quantum mechanical tunneling effect is considered. The dashed lines are classical reaction rates.

Further, the reaction rates of SF and PT processes were also investigated. One similarity between SF and PT processes is that the atoms involved in the motion of both are mainly hydrogen atoms. As the atom with the smallest mass, the quantum tunneling effect of the hydrogen atom cannot be ignored, so the quantum-corrected reaction rates were calculated. For the classical reaction rate can be obtained with the widely used Eyring equation,[39]

$$k_{classical}(T) = \frac{k_B T}{h} e^{-\frac{\Delta G(T)}{k_B T}} \quad (1)$$

where $k_B$ is Boltzmann's constant, and $h$ is Planck's constant. $\Delta G(T)$ denotes the free energy barriers at different temperatures. For the quantum-corrected reaction rate, the Eckart's tunneling correction method was used and the correction factors were calculated with the following function:[40]

$$\kappa(T) = \beta e^{\beta E^+} \int_{E_0}^{+\infty} P_{Eckart}(E - E_r) e^{-\beta E} \, dE \quad (2)$$

where $\beta = 1/k_B T$ and $k_B$ is Boltzmann's constant. $E^+$ represents the maximum energy for the potential barrier.

Interestingly, it is clear from Fig. 4(b) that the quantum tunneling effect leads to a crossover at a temperature of about 65 K, i.e., water ring rotation is more likely to occur when the temperature is greater than 65 K, and PT is more likely to dominate when the temperature is less than 65 K. Anyway, our research explicitly demonstrate that the DAE mechanism is highly temperature-dependent, exhibiting a dominance of PT process at low temperature and a dominance of SF processes at high temperature. Further, at a temperature of 300 K, the rate of overall rotation is $10^9$ s$^{-1}$ while the rate of PT is only the order of $10^{-1}$ s$^{-1}$, with a difference of 10 orders of magnitude. It suggests that DAE in water trimer under a conventional environment occurs mainly by flipping, rather than PT process previously thought. In this way, the discovery of SF-DAE has also changed the previous fundamental understanding of the HBNR mechanism in water trimer.

It can be known that the PT process is mainly the motion of H atoms along the O-H axis, while the SF process is mainly the motion of H atoms perpendicular to the O-H axis. This difference of quantum tunneling effect in SF and PT processes can be understood as a difference of quantum tunneling effect of H atoms in the two directions, as reported in the past.[41] To further explore the contribution of the two processes for different H atoms, we further discussed the effect of different D-atom substitution types on the two pathways. The quantum tunneling effect can be quantified using a reaction path described by IRC and the reduced mass corresponding to the frequency. Specifically, we calculated the tunneling probabilities ($P_{tunneling}$) for each reaction path according to the Wenzel-Kramers-Brillouin (WKB) approximation, as follows:[42–44]

$$P_{tunneling} = Exp\left[-\frac{2}{\hbar}\int_{x_1}^{x_2}\sqrt{2m(V(x)-E)}\,\mathrm{d}x\right] \qquad (3)$$

where V(x) are one-dimensional PESs obtained and $\hbar$ is reduced Planck constant. $x_1$ and $x_2$ represent the intersection points at different energies.

**Table II.** The vibration frequency (ν) of the transition state and the corresponding reduced mass, as well as the tunneling probability ($P_{tunneling}$) at the zero-point energy (ZPE) in sequential flipping (SF) and proton transfer (PT) processes based on the substitution of different atoms by deuterium (D).

| D-substituted atom | SF (ZPE = 0.067 eV) | | | PT (ZPE = 0.166 eV) | | |
|---|---|---|---|---|---|---|
| | ν (cm$^{-1}$) | Red. mass (a.u.) | $P_{tunneling}$ | ν (cm$^{-1}$) | Red. mass (a.u.) | $P_{tunneling}$ |
| No atoms | -139.08 | 1.062 | 6.31×10$^{-71}$ | -1810.56 | 1.026 | 4.61×10$^{-47}$ |
| 2 | -122.79 | 1.376 | 1.22×10$^{-80}$ | -1637.85 | 1.187 | 1.45×10$^{-50}$ |
| 6 | -122.79 | 1.376 | 1.22×10$^{-80}$ | -1633.36 | 1.193 | 1.12×10$^{-50}$ |
| 8 | -137.54 | 1.085 | 1.12×10$^{-71}$ | -1637.85 | 1.187 | 1.45×10$^{-50}$ |
| 9 | -137.54 | 1.085 | 1.12×10$^{-71}$ | -1809.36 | 1.027 | 4.47×10$^{-47}$ |
| 2, 6, 8 | -110.45 | 1.729 | 2.79×10$^{-90}$ | -1298.42 | 2.053 | 2.98×10$^{-66}$ |
| 2, 6, 8, 9 | -109.71 | 1.751 | 7.54×10$^{-91}$ | -1296.80 | 2.061 | 2.22×10$^{-66}$ |
| All H atoms | -100.65 | 2.160 | 7.71×10$^{-101}$ | -1292.85 | 2.081 | 1.06×10$^{-66}$ |

As can be seen in Table II, when the H atom is replaced by D, the frequency of the TS decreases and the mass corresponding to the reaction increases, leading to a decrease in the $P_{tunneling}$. This effect is more pronounced when the substituted atom is the main participant in the reaction process. It can be seen that atoms 2, 6, 8 and 9 are all involved in the SF process, where atoms 2 and 6 are equivalent and atoms 8 and 9 are equivalent. For the PT process, the three hydrogen atoms 2, 6 and 8 in the hydrogen bonding network of the water ring plane have similar effects. Specifically, atoms 2 and 8 are equivalent but have a slight difference for atom 6, which is due to the difference in the states in which the three H atoms are in the water trimer. Depending on the number of terminal hydrogens on the same side of the O-O-O plane in the initial Structure, water molecules H$^3$-O$^1$-H$^2$ and H$^8$-O$^7$-H$^9$ is referred to as the "majority" and water molecule H$^5$-O$^4$-H$^6$ is referred to as the "minority". The isotope effect amplifies the difference in vibration of water molecules in different states, which can also explain three small red-shifts (from 1201, 1639 and 1653 to 1203, 1641 and 1659 cm$^{-1}$) caused by the SF process in Fig. 2(a).

The PT process has a shorter reaction path and higher zero-point vibrational energy, which results in a higher probability of quantum tunneling for the same isotopic substitution conditions. When the temperature is high, the reaction occurs in a manner dominated by thermal motion. According to thermodynamics, the thermal motion is mainly related to the height of the energy barrier, which leads to SF processes being significantly faster than PT processes. When thermodynamic effects are no longer evident at low temperatures, this difference in quantum tunneling will be revealed and result in an intersection of reaction rate between the SF and PT processes at 65 K.

Based on the TST, the tunneling splitting of the reaction can be roughly inferred from the tunneling probability at the zero-point vibrational energy as well as the vibrational frequency.[17,45] From Table 2, it can be inferred that the tunneling splitting of the SF process will be much smaller than that of the PT process (1.9 × 10$^{-5}$ cm$^{-1}$ by large curvature tunneling effects and 8.2 × 10$^{-8}$ cm$^{-1}$ by small curvature tunneling effects),[17] which means that it is under the experimentally observed limit of about 1 MHz (3 × 10$^{-5}$ cm$^{-1}$).

## Conclusions

In summary, we have found an unexplored flip-dependent DAE mechanism in water trimer called SF process. This SF process has a smaller potential barrier than the collective rotation process, which is more in line with the Saykally's conjecture of a low potential barrier flipping process that can lead to DAE. The reaction pathway indicate that the SF process does not undergo O-H bond breakage, while the PT process undergoes three O-H bond breakings. This was further confirmed by IR and Raman spectra based on isotopic effects. Besides, the VCD and ECD spectra show that the products of the two processes differ in chirality. Interestingly, the reaction rate analysis shows a crossover point between the SF and PT processes due to the quantum tunneling effect. The SF process dominates at a temperature above 65 K, which suggested that the nature of what causes DAE to occur in most cases may not be the PT process that we previously thought but SF process. Our findings significantly enrich the understanding of the DAE mechanism in water trimer, which provides a piece of the jigsaw that has been sought to the HBNR mechanism.

## Author contributions

X. Yang performed the theoretical simulations, Z. Wang supervised the work. X. Yang, R. Liu, R. Xu and Z. Wang discussed the results. X. Yang and Z. Wang wrote the article

## Conflicts of interest

There are no conflicts to declare.

## Acknowledgements

The authors would like to acknowledge Dr. Depeng Zhang and Dr. Le Jin for constructive discussions. This work was supported by the 2020-JCJQ project (GFJQ2126-007) and the National Natural Science Foundation of China (grant number 11974136). Z. Wang also acknowledges the assistance of the High-Performance Computing Center of Jilin University and National Supercomputing Center in Shanghai.